\documentclass[reprint,aps,prd,onecolumn,notitlepage,nofootinbib,showpacs,preprintnumbers,superscriptaddress]{revtex4-1}
\usepackage[T1]{fontenc}
\usepackage[utf8]{inputenc}
\usepackage{amsmath}
\usepackage{amssymb}
\usepackage{amsfonts}
\usepackage{latexsym}
\usepackage{graphicx}
\usepackage{dcolumn}
\usepackage{bm}
\usepackage{empheq}
\usepackage{color}
\usepackage{hyperref}

\newcommand{\eq}[1]{\begin{equation}#1\end{equation}}

\newcommand{\ea}[1]{\begin{equation}\begin{aligned}#1\end{aligned}\end{equation}}

\newcommand{\lrp}[1]{\left( #1 \right)}  
\newcommand{\lrsb}[1]{\left[ #1 \right]}  
\newcommand{\lrab}[1]{\left\langle #1 \right\rangle}  

\def\rd{\partial}
\def\vk{\bm{k}}

\definecolor{gxhighlight}{rgb}{1,1,0.4}

\def\dep{\delta\phi}
\def\dop{\dot{\phi}}

\def\doq{\dot{Q}}

\begin{document}

\title{Primordial perturbations and non-Gaussianities in Ho\v{r}ava-Lifshitz gravity}

\author{Xian Gao}
\email{gaoxian@mail.sysu.edu.cn}
\affiliation{School of Physics and Astronomy, Sun Yat-sen University}

\date{\today}



\begin{abstract}
    We investigate primordial perturbations and non-gaussianities in the Ho\v{r}ava-Lifshitz theory of gravitation. In the UV limit, the scalar perturbation in the Ho\v{r}ava theory is naturally scale-invariant, ignoring the details of the expansion of the Universe. One may thus relax the exponential inflation and the slow-roll conditions for the inflaton field. As a result, it is possible that the primordial non-gaussianities, which are ``slow-roll suppressed'' in the standard scenarios, become large. We calculate the non-gaussianities from the bispectrum of the perturbation and find that the equilateral-type non-gaussianity is of the order of unity, while the local-type non-gaussianity remains small, as in the usual single-field slow-roll inflation model in general relativity. Our result is a new constraint on the the Ho\v{r}ava-Lifshitz gravity.
    \end{abstract}

\maketitle 

\section{Introduction}

A renormalizable theory of gravity was proposed by
Ho\v{r}ava \cite{Horava:2009uw,Horava:2008ih,Horava:2008jf}. This
theory reduces to Einstein's general relativity (GR) for large
scales, and may be a candidate for the UV completion of
general relativity. This theory is renormalizable in the sense that
the effective coupling constant is dimensionless in UV.  The
essential point of this theory is the anisotropic scaling of
temporal and spatial coordinates with dynamical critical exponent
$z$,
    \eq{
        t\rightarrow \ell^z t\,,\qquad x^i\rightarrow \ell x^i
        \,,\qquad\qquad (z\geq 1)\,.
    }
In $3+1$ spacetime dimension, the Ho\v{r}ava theory has an
ultraviolet fixed point with $z=3$. Since the Ho\v{r}ava theory is
analogue to the scalar-field model studied by Lifshitz, in which the full
Lorentz symmetry emerges only at the IR fixed point, the Ho\v{r}ava
theory is also called the Ho\v{r}ava-Lifshitz theory. Because of this
anisotropic scaling, time plays a privileged role in the Ho\v{r}ava
theory. In other words, spacetime has a codimension-one
foliation structure, which leaves the foliation  hypersurfaces of
constant time. Thus, contrary to GR, full diffeomorphism
invariance is abandoned, and only a subset (of the form of
local Galilean invariance) is kept. More precisely, the theory is
invariant under the foliation-preserving diffeomorphism defined by
    \eq{
        t\rightarrow \tilde{t}(t)\,,\qquad x^i \rightarrow
        \tilde{x}^i(t,x^i) \,.
    }
In the infrared (IR), due to a deformation by lower
dimensional operators, the theory flows to $z = 1$, corresponding to
the standard relativistic scale invariance under which the full
deffeomorphism, and thus general relativity, is recovered.
Non-relativistic scaling allows for many non-trivial scaling
theories in dimensions $D > 2$. The Ho\v{r}ava-Lifshitz theory allows a
theory of gravitation that is scale-invariant in UV, while the
standard GR with full diffeomorphism emerges at the IR fixed point.

The Ho\v{r}ava-Lifshitz gravity theory has been intensively
investigated
\cite{Horava:2009if,Volovich:2009yh,Jenkins:2009un,Kiritsis:2009sh,Takahashi:2009wc,Calcagni:2009ar,Mukohyama:2009gg,Brandenberger:2009yt,Lu:2009em,Piao:2009ax,Nikolic:2009jg,Nastase:2009nk,Cai:2009pe,Cai:2009ar,Chen:2009ka} (see also \cite{Mukohyama:2010xz,Wang:2017brl} for reviews and more references therein).
In particular, cosmology in the Ho\v{r}va theory has been studied in
\cite{Kiritsis:2009sh,Takahashi:2009wc,Calcagni:2009ar,Mukohyama:2009gg,Brandenberger:2009yt,Lu:2009em,Piao:2009ax}.
Homogeneous vacuum solutions in this theory were obtained in
\cite{Lu:2009em}, and scalar and tensor perturbations were studied
in \cite{Takahashi:2009wc,Mukohyama:2009gg}. In \cite{Kiritsis:2009sh,Calcagni:2009ar}, the cosmological evolution in
Ho\v{r}ava gravity with scalar-field was extensively studied, and
the matter bounce scenario in the Ho\v{r}ava theory was investigated by
Brandenberger \cite{Brandenberger:2009yt}.

As was pointed out in \cite{Kiritsis:2009sh}, the Ho\v{r}ava theory has at
least two important properties. The first is its UV
renormalizability, while the second is more interesting for
cosmology. The fact that the speed of light is diverging in UV
implies that exponential inflation is not necessary for
solving the horizon problem. Moreover, the short distance structure of
perturbations in the Ho\v{r}ava-Lifshitz theory is different from the
standard inflation in GR. In particular, in the UV limit, the scalar field
perturbation is essentially scale-invariant and is insensitive to
the expansion rate of the Universe, as has been addressed in
\cite{Kiritsis:2009sh,Takahashi:2009wc,Calcagni:2009ar,Mukohyama:2009gg}.
The key point is that the UV renormalizability indicates that the
Lagrangian for the non-relativistic scalar field should contain up
to six spatial derivatives. Thus, in the UV limit, the dispersion
relation is $\omega^2 \sim k^6$, which is contrary to $\omega^2 \sim
k^2 $ in standard GR. This phenomenon causes different $k$-dependence
of the two-point correlation function and thus the scalar
perturbation in Ho\v{r}ava gravity is naturally scale-invariant in the
UV limit.

In this paper, we extend the previous works on cosmological perturbation
theory in Ho\v{r}ava gravity, including non-gaussianities.
The 6-parameter $\Lambda$CDM model provides an accurate description of the Universe \cite{Aghanim:2018eyx}. In particular, the primordial perturbations are assumed to be Gaussian. 
Deviation from the Gaussian distribution, i.e., primordial non-gaussianity, has not been observed.
This puts a strict constraint on any model of the early Universe.
The non-Gaussian features of Ho\v{r}ava gravity have not been studied in detail, except in \cite{Huang:2012ss} and in \cite{Huang:2013epa}, which investigated the non-gaussianities of the scalar field and of the gravitational waves in Ho\v{r}ava gravity, respectively.
Actually, one of the essential differences of Ho\v{r}ava gravity from
Einstein's general relativity is that it contains quadratic curvature
terms in the theory. Moreover, the
foliation-preserving diffeomorphism does not allow, unfortunately, to choose a spatially-flat
gauge as in GR. Thus, in general, the perturbation theory in
Ho\v{r}ava gravity in quite involved. On the other hand, the number of dynamical degrees of freedom in the spatial metric is 3 in Ho\v{r}ava gravity
(contrary to 2 in GR), with 2 tensor degrees of
freedom, as usual, and an additional scalar dynamical degree
of freedom (see also \cite{Gao:2014soa,Gao:2014fra,Gao:2018znj} for the general framework of spatially covariant theories of gravity).

In this work, we focus on the perturbation of the scalar field. Thus,
for simplicity, we neglect the spatial metric perturbation. We
introduce a scalar field, following the strategy in
\cite{Kiritsis:2009sh,Calcagni:2009ar}. We pay special attention to the
non-gaussianities in this scalar field model in Ho\v{r}ava gravity.
The basic idea is that, as has been addressed before, the divergence
of the speed of light and the scale-invariance of the scalar
perturbation in Ho\v{r}ava gravity indicate that there is no need to
assume an exponential expansion of the Universe. Moreover, the
traditional slow-roll conditions are not necessary. While it is
well-known that in slow-roll inflationary models non-gaussianity is
suppressed by slow-roll parameters \cite{Maldacena:2002vr}, and thus
too small to be detected (see e.g. \cite{Bartolo:2004if} for a 
review of non-gaussianities in cosmological perturbations. Various
models have been investigated to generate large non-gaussianities by
introducing more complicated kinetic terms
\cite{Seery:2005wm,Chen:2006nt,Huang:2006eha,Arroja:2008ga} or more
fields
\cite{Seery:2005gb,Langlois:2008qf,Langlois:2008wt,Arroja:2008yy,Gao:2008dt,Gao:2009gd}).
However, in Ho\v{r}ava gravity, there are no slow-roll conditions,
and thus the ``slow-roll suppressed'' non-gaussianities can become large. In this work, we focus on the non-gaussianity from
the bispectrum, which is defined from the three-point correlation function of
the perturbation. We find that the equilateral-type non-gaussianity
is roughly of the order of unity, while the local-type non-gaussianity remains
very small, as in the usual single-field slow-roll inflation in GR.

The paper is organized as follows. In Section 2, we briefly review
the Ho\v{r}ava gravity and set out our conventions. In Section 3, we
couple the scalar field to Ho\v{r}ava gravity, and describe the
cosmological evolution of the Ho\v{r}va gravity/scalar matter
system. In Section 4, we calculate the scalar field perturbation,
including the gravity perturbations. We get the full second-order
perturbation action, which reduces to the standard result in the IR
limit. In the UV limit, the scalar perturbation is essentially
scale-invariant. In Section 5, we calculate the non-gaussianities.
Finally, we make a conclusion and discuss several related issues.

\section{Brief Review of Ho\v{r}ava-Lifshitz Gravity}

In this section we briefly review the Ho\v{r}ava-Lifshitz
theory\cite{Horava:2009uw}. The dynamical variables in
Ho\v{r}ava-Lifshitz gravity are the spatial scalar $N$, spatial vector
$N_i$ and spatial metric $g_{ij}$. This is similar to the ADM
formalism of the metric in standard general relativity, while in
Ho\v{r}ava gravity, $N$, $N_i$ and $g_{ij}$ are related to the
space-time metric as
    \eq{
        ds^2 = -N^2 c^2 dt^2 + g_{ij} \lrp{ dx^i +N^i dt } \lrp{dx^j + N^j dt}
        \,,
    }
where $c$ is the speed of light.

The action of Ho\v{r}ava-Lifshitz gravity contains a ``kinetic'' part and
a ``potential'' part, \eq{
    S= S_{K} +S_{V} \,,
} with
    \eq{{\label{g_k_action}}
    S_K =  \frac{2}{\kappa^2} \int dt d^3x \sqrt{g} N
   \left( K_{ij}K^{ij}-\lambda K^2 \right)   \ ,
    }
where
    \[
    K_{ij}= \frac{1}{2N} \lrp{\dot{g}_{ij}- \nabla_i N_{j} - \nabla_j N_{i} } \, ,
    \]
is the extrinsic curvature and $K= g^{ij}K_{ij}$. The potential
terms are given in the ``detailed-balance'' form
    \ea{
    S_V & = \int dt d^3x \sqrt{g}  N  \lrsb{ -\frac{\kappa^2}{2 w^4}C_{ij}C^{ij}
     +\frac{\kappa^2\mu}{2 w^2}\epsilon^{ijk}R_{il} \nabla_j R^{l}_{k}
       -\frac{\kappa^2\mu^2}{8}R_{ij}R^{ij} +\frac{\kappa^2\mu^2}{8(1-3\lambda)}\left(\frac{1-4\lambda}{4}R^2+\Lambda R-3\Lambda ^2\right)
     }
     \,,
    }
where $C_{ij}$ is the Cotton tentsor defined by
    \eq{
        C^{ij}= \frac{\epsilon^{ikl}}{\sqrt{g}} \nabla_k \lrp{
        R^{j}_{l}-\frac{1}{4}R\delta^{j}_{l} }
        \,.
    }
Note that in (\ref{g_k_action}) $\lambda$ is a dimensionless
coupling of the theory, and therefore runs.

As mentioned in the introduction, the essential point of the Ho\v{r}ava
theory is the anisotropic scaling of temporal and spatial
coordinates: $t\rightarrow \ell^z t$ and $ x^i\rightarrow \ell x^i$.
The classical scaling dimensions of various quantities in the Ho\v{r}ava
theory are summarized in Tab.\ref{tab_dims}.
\begin{table}[h]
\centering
    \begin{tabular}{|r|c|c|c|c|c|c|c|c|c|c|c|}
      \hline
      & $[t]$ & $[x^i]$ & $[c]$ & $[\kappa]$ & $[N]$ & $[N_i]$ & $[g_{ij}]$ & $[\Delta]$ & $[\rd_t]$ & $[\rd_i]$ & $[\phi]$ \\
       \hline
       dimension &$-z$ & $-1$ & $z-1$ & $\frac{z-3}{2}$ & $0$ & $z-1$ & $0$ & $2$ & $z$ & $1$ & $\frac{3-z}{2}$  \\
      \hline
    \end{tabular}
    \caption{\label{tab_dims} Summary of the classical scaling dimensions of various quantities in the Ho\v{r}ava-Lifshitz theory.}
\end{table}

\section{Cosmology with Scalar Field Matter}

\subsection{Coupling the scalar field to Ho\v{r}ava-Lifshitz gravity}

In this work, we couple the scalar-field matter with Ho\v{r}ava-Lifshitz
theory following the strategy in
\cite{Kiritsis:2009sh,Calcagni:2009ar}. The general structure of the
action of the scalar-field matter and Ho\v{r}ava-Lifshitz gravity contains
two parts: a quadratic kinetic term with the foliation-preserving
diffeomorphisms and a potential term:
    \eq{{\label{scalar_action}}
        S^{\phi} = \int dt d^3x\, \sqrt{g}N
        \lrsb{ \frac{1}{2N^2} \lrp{ \dop - N^i \rd_i \phi }^2  + F(\phi,\rd_i\phi,g_{ij})
        }\,.
    }
The ``potential term'' is
    \eq{{\label{scalar_f}}
        F  = -V(\phi) + g_1 \xi_1 +
        g_{11} \xi_1^2 + g_{111} \xi_1^3 + g_2
        \xi_2 + g_{12} \xi_1 \xi_2 + g_3 \xi_3 \,,
    }
where $\xi_{i}$ and their properties in UV/IR are summarized in
Tab.\ref{tab_op}.
\begin{table}[h]
\centering
\begin{tabular}{|r|c|c|c|c|}
  \hline
  & $\mathcal{O}$   &scaling dim $[\mathcal{O}]$ & $z=3$, UV fixed point & $z=1$, IR fixed point  \\
  \hline
  & $\dop^2$ &  $z+3$ & marginal & marginal \\
  $\xi_{1}$ & $\rd^i\phi \rd_i\phi$ &$5-z$ & relevant & marginal \\
  $\xi_{1}^2$ &  $\lrp{ \rd^i\phi \rd_i\phi }^2$  & $10-2z$ & relevant & irrelevant \\
  $\xi_{1}^3$ & $\lrp{ \rd^i\phi \rd_i\phi }^3$  &$15-3z$ & marginal & irrelevant \\
  $\xi_{2}$ & $ \lrp{\Delta\phi }^2 $  & $7-z$ & relevant & irrelevant \\
  $\xi_{1} \xi_2$ & $ \lrp{ \rd^i \phi \rd_i\phi} \lrp{ \Delta \phi }^2 $ & $12-2z$ & marginal & irrelevant \\
  $\xi_{3}$ & $ (\Delta \phi)(\Delta^2\phi) $ & $9-z$ & marginal & irrelevant \\
  \hline
\end{tabular}
\caption{\label{tab_op} Summary of the operators in the
non-relativistic scalar field action and their properties under the
renormalization group flow from $z=3$ (UV) to $z=1$ (IR).}
\end{table}
In Tab.\ref{tab_op}, $\Delta = g^{ij}\nabla_i\nabla_j$ is the
spatial Laplacian, and $g_{1},g_{11}$ etc. can be constant, or in general they can be 
functions of $\phi$. We assume $g_3>0$ in order to guarantee the
stability of the perturbation in UV.

 In the UV limit ($z=3$ fixed point), $\xi_1^3$, $\xi_1\xi_2$ and $\xi_3$  dominate. Thus, in UV the
scalar field action takes the form
    \eq{{\label{scalar_action_UV}}
        S^{\phi}_{\textrm{UV}} = \int dt d^3x\, a^3 N
        \lrsb{ \frac{1}{2N^2} \dop^2  + g_{111} \lrp{\rd^i \phi \rd_i \phi
        }^3 + g_{12} \lrp{\rd^i \phi \rd_i \phi} \lrp{\Delta \phi}^2
        + g_3 \lrp{ \Delta \phi } \lrp{ \Delta^2\phi}
        }\,.
    }

\subsection{Equations of motion}

The full equations of motion for $N$, $N_i$ and $g_{ij}$ have been
derived in \cite{Kiritsis:2009sh,Calcagni:2009ar,Lu:2009em}. For our purpose we focus on the equations of motion for $N$
and $N_i$, which we write here for later convenience :
\ea{{\label{constraint}}
    0 &= -\frac{2}{\kappa^2} \lrp{ K_{ij}K^{ij} - \lambda K^2 } -\frac{\kappa^2}{2 w^4}C_{ij}C^{ij}
     +\frac{\kappa^2\mu}{2 w^2}\epsilon^{ijk}R_{il} \nabla_j R^{l}_{k}
       -\frac{\kappa^2\mu^2}{8}R_{ij}R^{ij} \\
       &\qquad\qquad + \frac{\kappa^2\mu^2}{8(1-3\lambda)}\left(\frac{1-4\lambda}{4}R^2+\Lambda R-3\Lambda
       ^2\right) -\frac{1}{2N^2} \lrp{ \dop -N^i\rd_i \phi }^2 + F
       \,,\\
    0 &= \frac{4}{\kappa^2} \nabla_j\lrp{ K^j_i -\lambda K \delta^j_i } -
    \frac{1}{N} \lrp{ \dop - N^i \rd_i \phi } \rd_i \phi \,.
}

\subsection{Cosmological Evolution}

We now consider the cosmological background evolution in
Ho\v{r}ava-Lifshitz gravity. We assume the background to be
homogeneous and isotropic, and use the residual invariance under
time re-parametrization to set $N=1$. We focus on the flat
3-dimensional case. The background values are
    \eq{
        N=1\,,\qquad N_i=0\,,\qquad g_{ij}=a^2(t) \delta_{ij} \,,\qquad \phi_0 =\phi_0 (t) \,.
    }
In this background, the action in the gravity sector is significantly
simplified. In particular, $R_{ij} = C_{ij} =0 $ and the spatial covariant
derivatives mostly vanish.

The equation of motion for $N$ to the 0-th order gives
    \eq{
        3(3\lambda - 1) H^2 \alpha +\sigma - \frac{\dop_0^2}{2} -  V(\phi_0) =0
        \,,
    }
where we have denoted
    \eq{
        \alpha = \frac{2}{\kappa^2} \,,\qquad \sigma = -\frac{3\kappa^2\mu^2\Lambda^3}{8(1-3\lambda)} \,,
    }
and $H\equiv \dot{a}/a$ is the familiar Hubble parameter.
The equation of motion for $g_{ij}$ gives
    \eq{
        2(3\lambda - 1) \alpha \lrp{ \dot{H} + 3H^2/2 } +
        \frac{\dop_0^2}{2} - V(\phi_0) =0 \,.
    }
The equation of motion for the scalar field is
    \eq{
        \ddot{\phi}_0 + 3H \dop_0 + \frac{V'_0}{2} = 0\,.
    }

\section{Cosmological Perturbation}

We now consider cosmological perturbation in Ho\v{r}ava gravity
coupled to scalar-field matter.

As has been addressed before, the
action of Ho\v{r}ava gravity is complicated due to the
quadratic terms in spatial curvature. We recall that in GR, one
can choose various gauges to simplify the calculations. However, the
case is different in Ho\v{r}ava gravity (see Appendix A for a
 discussion of gauge transformation and gauge choice in
Ho\v{r}ava theory). Since the ``foliation-preserving'' diffeomorphism
is only a subset of the full diffeomorphism in GR, 
one may expect in general that there are less gauge modes and more physical
modes\footnote{As argued in the original proposal by Ho\v{r}ava
\cite{Horava:2009uw}, the dynamical degrees of freedom in the
spatial metric $g_{ij}$ is 3 (with contrary to 2 in GR) when
$\lambda \neq 1 \;\textrm{and}\; 1/3$, which contains 2 usual tensor
degrees of freedom and \emph{one additional scalar degree of
freedom} (see also similar arguments based on cosmological context
and more general analysis in ). }. Thus, the perturbation theory in
Ho\v{r}ava-Lifshitz theory is very involved, but also interesting.

We consider the scalar-field perturbation by neglecting the spatial metric perturbation. We assume that the background scalar field
is homogeneous $\phi_0 = \phi_0(t)$.

As we focus on the scalar perturbation, we write (in
the background $N=1$ gauge){\footnote{In Ho\v{r}ava's original
formulation of the theory, $N$ was assumed to be a function of time
only, $N=N(t)$. Here in this work, we relax this restriction to
assume $N$ to be function of both temporal and spatial
coordinates.}}
    \ea{
        N &\equiv 1 + \alpha_1 + \cdots\,,\\
        N_i &\equiv \rd_i \beta_1 +\theta_{1i} +\cdots \,.
    }
Here, the subscript ``1'' denotes the first-order in $\dep \equiv \delta
\phi$. The constraints in Eq. (\ref{constraint}) become
\ea{{\label{constraint_sf}}
    0 &= -\frac{2}{\kappa^2} \lrp{ K_{ij}K^{ij} - \lambda K^2 } +\sigma -\frac{1}{2N^2} \lrp{ \dop -N^i\rd_i \phi }^2 + F
       \,,\\
    0 &= \frac{4}{\kappa^2} \nabla_j\lrp{ K^j_i -\lambda K \delta^j_i } -
    \frac{1}{N} \lrp{ \dop - N^i \rd_i \phi } \rd_i \phi \,.
} Solving  equations (\ref{constraint_sf}) up to the first-order of
$\dep$, we get {\footnote{At this point, it is useful to compare
(\ref{constraint_solutions}) with the standard results in
perturbation theory in GR. It can be seen directly that in the IR
fixed point where $z=1$, if we choose $\lambda=1$,
(\ref{constraint_solutions}) reduces to the previous well-known
results in GR.}}
    \ea{{\label{constraint_solutions}}
        \alpha_1 &= \frac{(-1+\lambda ) \dot{Q} \dot{\phi }_0+Q \left(H (-1+3 \lambda ) \dot{\phi }_0+(-1+\lambda ) V'\right)}{4 H^2 \alpha  (-1+3 \lambda )+(-1+\lambda ) \dot{\phi }_0^2} \,,\\
        \rd^2 \beta_1 &= a^2 \frac{2  H \alpha  (1-3 \lambda ) \dot{Q} \dot{\phi }_0+ Q \left(6 H^2 \alpha  (1-3 \lambda ) \dot{\phi }_0+\dot{\phi }_0^3+2 H \alpha  (1-3 \lambda ) V'\right)}{2 \alpha  \left(4 H^2 \alpha  (-1+3 \lambda )+(-1+\lambda ) \dot{\phi }_0^2\right)} \,,
    }
and $\theta_{1i}=0$, as usual in GR.  Note that when $\lambda \neq 1$, $\alpha_1$ also depends on
 $\doq$, which is different from GR, where $\alpha_1 \sim
 \dep$. It is useful to note that in the UV limit  $\alpha_1$ and $\beta_1$ become
    \ea{{\label{constraint_UV}}
        \alpha_1 &\simeq \frac{\sqrt{3} (\lambda-1 ) \dot{Q}}{H \sqrt{2 \alpha } (3 \lambda-1 )^{3/2}} \,,\\
        \rd^2 \beta_1 &\simeq -\frac{\sqrt{3} a^2 \dot{Q}}{\sqrt{2\alpha  (3 \lambda-1 )}} \,,
    }
where we have used the background equations of motion in the UV
limit.

\subsection{Linear perturbation}

As we are neglecting the spatial metric perturbation, the
gravity sector is greatly simplified
    \eq{
        S^g = \int dt d^3x a^3 N
   \lrsb{ \alpha \left( K_{ij}K^{ij}-\lambda K^2 \right) +\sigma}  \,.
    }

After a rather tedious but straightforward calculation, we get the
quadratic part of the action for the scalar perturbation $Q$:
\ea{{\label{full_2nd_action}}
    S_2[Q] =\int dtd^3x\, a^3 \lrsb{ \frac{\gamma}{2} \doq^2 + \omega \doq Q +
    m
    Q^2 + g_1 \rd^i Q \rd_i Q + g_2 (\Delta Q)^2 + g_3 (\Delta Q) (\Delta^2 Q) }\,,
    } where
    \ea{{\label{coefficients}}
        \gamma &= \frac{H^2 \alpha  (1-3 \lambda )^2 \left[ H^2 \alpha  (7+3 \lambda  (3 \lambda-4 ))-(\lambda -1) (V_0-\sigma) \right]}{ \left[ H^2 \alpha  (1-3 \lambda )^2-(\lambda-1)
        (V_0-\sigma) \right]^2} \,,\\
        \omega &= - \frac{ 2 H^2 \alpha  (3 \lambda -1) - (\lambda-1 ) (V_0-\sigma) }{ 2 \left[H^2 \alpha  (1-3 \lambda )^2-(\lambda-1 )
        (V_0-\sigma)\right]^2 }  \left[ 6 H^3 \alpha  (1-3 \lambda )^2+H (2-6 \lambda ) (V_0-\sigma) \right. \\
        &\qquad\qquad\qquad\qquad \left. + (\lambda-1 ) \sqrt{6 H^2 \alpha  (-1+3 \lambda )-2 (V_0-\sigma) } V'\right] \,,\\
        m &= \frac{1}{4 \alpha  \left(H^2 \alpha  (1-3 \lambda )^2-(\lambda -1) (V_0-\sigma)\right)^2 }
        \left\{ 18 H^6 \alpha ^3 (1-3 \lambda )^4+(2-2 \lambda ) (V_0-\sigma)^3-2 H^2 \alpha ^2 (\lambda-1 ) (3 \lambda -1) \left(V'\right)^2 \right. \\
        &\qquad + \sqrt{6 H^2 \alpha  (-1+3 \lambda )-2 (V_0-\sigma)} \left(-4 H^3 \alpha ^2 (1-3 \lambda )^2 V'+2 H \alpha  (\lambda-1 ) (3 \lambda -1) (V_0-\sigma) V'\right) \\
        &\qquad -2 H^4 \alpha ^3 (1-3 \lambda )^4 V''  +(V_0-\sigma) \left(-12 H^4 \alpha ^2 (3 \lambda-1 )^3+\alpha  (\lambda -1)^2 \left(V'\right)^2+4 H^2 \alpha ^2 (1-3 \lambda )^2 (\lambda -1) V''\right) \\
        &\qquad \left. + (V_0-\sigma)^2 \left(4 H^2 \alpha  (2+9 (-1+\lambda ) \lambda )-2 \alpha  (\lambda-1 )^2
        V''\right) \right\} \,.
    }
In (\ref{full_2nd_action}), $g_i \equiv g_i(\phi_0)$.
In deriving Eq. (\ref{full_2nd_action}), we have used the background
equations of motion. Moreover, no approximations were made in deriving Eq.
(\ref{full_2nd_action}) and thus it is exact. We can use Eq.
(\ref{full_2nd_action}) to analyze the behavior of perturbations
both in the IR or UV limits and in the interpolation era.

\subsubsection{IR limit}

Taking the IR limit of the full second-order perturbation action Eq.
(\ref{full_2nd_action}) and choosing $\lambda = 1$, we get
    \eq{{\label{2rd_IR}}
        S_{2}^{\textrm{IR}} = \int dtd^3x a^3 \lrsb{ \frac{1}{2} \doq^2 + g_1 \rd^i Q \rd_iQ - \frac{H\epsilon}{2\alpha} \doq Q + H^2\lrp{ \frac{\epsilon^2}{2\alpha^2} -\frac{ \sqrt{\epsilon \eta_1}}{2 \alpha} -\frac{\eta_2}{2} } Q^2  } \,,
    }
where we have defined the dimensionless parameters
    \ea{
        \epsilon &= \frac{\dop^2}{2H^2} \,,\\
        \eta_1 &= \frac{1}{2}\lrp{\frac{V'}{H^2}}^2\,,\\
        \eta_2 &= \frac{V'}{H^2} \,.
        }
If we further set $\alpha= \frac{1}{2} $ and choose $g_1 =
-\frac{1}{2}$, Eq. (\ref{2rd_IR}) reduces to the familiar result in the
perturbation theory in GR. Especially, the perturbation $Q$ is
scale-invariant when the expansion of the Universe is exponential,
and thus with an approximately constant Hubble parameter $H\approx
\textrm{const}$.

\subsection{Scale-invariant spectrum in Ho\v{r}ava-Lifshitz era}

We now focus on the behavior of the perturbation theory in the
UV-limit, where $\doq^2$ and $Q\Delta^3Q$ terms dominate. The
perturbation action Eq. (\ref{full_2nd_action}) becomes rather simple in
the UV limit,
    \ea{{\label{2rd_action}}
        S_2[Q] &= \int dtd^3x\, a^3 \lrp{ \frac{\gamma}{2} \dot{Q}^2 + g_3\, \Delta Q \Delta^2Q  } \,,
    }
with
    \ea{
        \gamma &= \frac{ 7+3 \lambda  (3 \lambda-4) }{ (3 \lambda -1)^2}
        \,.
    }
Note that $\gamma$ is now constant. It is convenient to use a new
variable $u$, defined as $u \equiv a \sqrt{\gamma} Q$.
After changing into conformal time $\eta$, defined by $dt = a d\eta$,
and going into Fourier space, the second-order perturbation action
reads
    \eq{
        S_2 = \int d\eta \frac{d^3k}{(2\pi)^3} \lrsb{ \frac{1}{2} \lrp{ u'_{\vk} - \mathcal{H}u_{\vk} }\lrp{ u'_{-\vk} - \mathcal{H}u_{-\vk} } +\frac{g_3 k^6}{\gamma a^4}  u_{\vk} u_{-\vk}  }  \,.
    }

The equation of motion for the perturbation reads
    \eq{{\label{eom}}
        u''_{k} + \lrp{ \frac{g_3}{\gamma}\frac{k^6}{a^4} - \frac{a''}{a}
        } u_{k} = 0\,.
    }
Here, we assume for simplicity that $g_3$ is approximately constant. The mode
function is
    \eq{{\label{mode_function}}
        u_k(\eta) =  \frac{ \lrp{\gamma/g_3 }^{\frac{1}{4}} }{ \sqrt{2 k^3 }}\, a(\eta) \, \exp\lrp{-i \sqrt{ \frac{g_3}{\gamma} }\, k^3 \int^{\eta} \frac{d\eta'}{a^2(\eta')} }\,.
    }
The mode function is chosen such that it satisfies the Wronskian
normalization condition:
    \eq{
        u_{k}'(\eta)u_{k}^{\ast}(\eta)- u_{k}'^{\ast}(\eta)u_{k}(\eta)=
        -i \,.
    }
Moreover, the short-time behavior of the mode function Eq.
(\ref{mode_function}) is analogue to that of a positive-frequency
oscillator.

The tree-level two-point correlation function of $Q(\vk,\eta)$ is
\eq{{\label{2pf}}
    \lrab{ Q(\vk_1,\eta_1) Q(\vk_2,\eta_2) } = (2\pi)^3 \delta^3(\vk_1 +
    \vk_2) \frac{1}{ \sqrt{\gamma g_3}\, 2 k_1^3} \exp\lrp{-i \sqrt{ \frac{g_3}{\gamma} }\, k_1^3 \int^{\eta_1}_{\eta_2} \frac{d\eta'}{a^2(\eta')} }  \,,
} and thus the power spectrum of $Q$ is given by
    \eq{{\label{powersp}}
        \lrab{ Q(\vk_1,\eta_{\ast}) Q(\vk_2,\eta_{\ast}) } =
        (2\pi)^3 \delta^3(\vk_1 + \vk_2) P(k_1)\,,
    }
with
    \eq{
        P(k_1) = \frac{1}{ \sqrt{\gamma g_3}\, 2 k_1^3} \,.
    }
The so-called dimensionless power spectrum of $Q$ is
    \eq{{\label{spectrum}}
        \Delta^2(k) \equiv \frac{k^3}{2\pi^2} P(k) = \frac{1}{(2\pi)^2} \frac{1}{\sqrt{ \gamma g_3 } } \equiv \textrm{const.} \,.
    }
 The power spectrum of the
scalar perturbation is naturally scale-invariant in the UV limit,
ignoring the details of the expansion of the Universe. This feature
is contrary to that in GR, where a nearly constant Hubble expansion
rate $H$ is needed to guarantee the scale-invariance of the spectrum.
Due to this fact, there is no need to take any ``slow-roll''-type
conditions for the scalar field.

We would like to make some comments here. The crucial picture of the
standard inflation is that quantum fluctuations are generated in the
subhorizon region ($k\gg aH$), and are then stretched to the cosmological
size and become classical ($k \ll aH$). The horizon-exiting point
corresponds to $k = aH$. Thus, the ``horizon-exiting'' process
exists only when $aH$ is an increasing function of time. If we
assume a power law inflation $a \propto t^p$, it requires $p>1$. This
is violated by the curvature, and is only satisfied when the equation of
state $w <-1/3$. This is why in the standard inflation model we need a
slow-rolling scalar field to mimic the cosmological constant and to
drive an exponentially expanding background. However, the
``horizon-exiting'' process occurs generically in the
Ho\v{r}ava-Lifshitz era for rather general cosmological backgrounds.
From Eq. (\ref{eom}), it is obvious that the perturbation stops
oscillating when
\[
    \frac{k^6}{a^6} \sim H^2 \,,
\]
and thus requires that $a^6H^2$ is an increasing function of time.
Obviously, if we assume a power law expansion $a \propto t^p$ , this
demands $p>1/3$. In terms of the comoving time $a \propto |\eta|^p$,
we need $p<1/2$. This condition can be satisfied by any matter
component with the equation of state $w <1$.

We get the scale-invariant power spectrum in the UV
limit from the equation of motion Eq. (\ref{eom}). While one may
consider the full equation of motion, the second-order
action Eq. (\ref{full_2nd_action}) can be used. In general, the equation of motion
for the perturbation has the following functional form:
    \eq{{\label{full_2rd_eom}}
        u''_k + \lrp{ c_1^2 k^2 + \lambda_1 \frac{ \ell^2 k^4}{a^2} + \lambda_2 \frac{ \ell^4 k^6}{a^4} - \frac{a''}{a} + m^2 a^2
        } u_k = 0 \,,
    }
where $c_1$, $\lambda_1$ and $\lambda_2$ are dimensionless
parameters, $m$ is the effective mass parameter, and $\ell$ is the
length scale of the whole theory. The functional form of the
dispersion relation in (\ref{full_2rd_eom}) has been intensively studied
in the investigation of trans-Planckian effects
\cite{Martin:2000xs,Brandenberger:2000wr,Starobinsky:2001kn,Martin:2002kt}, and
also in statistical anisotropy \cite{Gao:2009vi}. Although complicated,
it is  interesting and important to investigate Eq. (\ref{full_2rd_eom})
in order to understand  the behavior of the perturbation not only in
the UV limit, but also in the interpolation region between UV and IR.

\section{Non-gaussianities}

In this section, we investigate the non-gaussianities, which
characterize the \emph{interaction} of the perturbations.

\subsection{Bispectrum}

We focus on the third-order perturbation action and
the three-point correlation function of the perturbation $Q$. The
third-order action in the gravity sector is \eq{
    S^g_3 = \int dtd^3x\,a^3 \alpha \lrsb{ - \frac{\alpha_1}{a^4}\lrp{ \rd_i\rd_j \beta_1 \rd_i\rd_j \beta_1 - \lambda (\rd^2\beta_1)^2 } - 2(1-3\lambda) \frac{H}{a^2} \alpha_1^2 \rd^2\beta_1 - 3(1-3\lambda) H^2 \alpha_1^3
    } \,.
} and in the scalar field sector is \ea{
    S^{\phi}_3 &= \int dtd^3x\, a^3 \left\{  \frac{1}{2} \lrsb{ -2\doq \rd^i\beta_1 \rd_i Q - \alpha_1 \lrp{ \doq^2 - 2\dop \rd^i \beta_1 \rd_i Q } + 2 \dop \doq \alpha_1^2 - \dop^2 \alpha_1^3  }  \right. \\
    &\qquad\qquad\qquad\qquad -\frac{V'''}{6}Q^3 + g'_1 Q \rd^iQ\rd_i Q + g'_2
    Q(\Delta Q)^2 + g'_3 Q(\Delta Q)(\Delta^2 Q) \\
    &\qquad\qquad\qquad\qquad \left. +\alpha_1 \lrp{ -\frac{V''}{2}Q^2 + g_1 \rd^iQ\rd_i Q + g_2 (\Delta Q)^2 + g_3 (\Delta Q)(\Delta^2 Q)
    } \right\} \,,
} where $\alpha_1$ and $\beta_1$ are given in Eq.
(\ref{constraint_solutions}).
 We are interested in the UV behavior of the perturbation. In the
UV limit, after a straightforward calculation, the third-order
perturbation action reads,
    \eq{{\label{S3_UV}}
        S_3^{\textrm{UV}}[Q] = \int dt d^3x\, \frac{a^3}{H} \lrsb{ b_1 \dot{Q}^3  + b_2 \dot{Q} \Delta Q \Delta^2Q  + b_3 \doq \lrp{ \frac{\rd_i\rd_j}{\rd^2} \doq }^2  } \,.
    }
where
    \ea{
        b_1 &= \frac{\sqrt{\frac{3}{2}} (\lambda -1) (8+3 \lambda  (3 \lambda-5 )) }{2 \sqrt{\alpha } (3 \lambda
        -1)^{7/2}} \,,\\
        b_2 &= \frac{\sqrt{\frac{3}{2}}   (\lambda-1 )  g_3}{\sqrt{\alpha } (3 \lambda
        -1)^{3/2}} \,,\\
        b_3 &= - \frac{3\sqrt{3}(\lambda-1) }{2 \sqrt{2\alpha} (3\lambda-1)^{5/2}
        }\,,
    }
which are dimensionless constants (recall that we assume $g_3$ to be
approximately constant). In Eq. (\ref{S3_UV}), the formal operator
$\frac{\rd_i \rd_j}{\rd^2}$ should be understood in momentum space.
After changing into comoving time $\eta$ and into Fourier
space, we have
    \ea{{\label{s3_UV_fourier}}
        S_3^{\textrm{UV}} &= \int d\eta \prod_{i=1}^3 \frac{d^3 k_i}{(2\pi)^3}\, (2\pi)^3 \delta^3(\vk_{123}) \left[ \frac{a}{H} \lrp{b_1 + b_3 (\hat{\vk}_2 \cdot \hat{\vk}_3)^2} Q'(\vk_1,\eta)Q'(\vk_2,\eta)Q'(\vk_3,\eta) \right.\\
        &\qquad\qquad\qquad\qquad\qquad\qquad \left. - \frac{b_2}{a^3H} k_2^2 k_3^4 Q'(\vk_1,\eta)Q(\vk_2,\eta)Q(\vk_3,\eta)
        \right]\,,
    }
where we denote $\vk_{123} \equiv \vk_1 + \vk_2 + \vk_3$.

The three-point correlation function in cosmological context is
evaluated in the so-called ``in-in'' formalism
 \eq{
    \lrab{ Q(\vk_1,\eta_{\ast}) Q(\vk_2,\eta_{\ast}) Q(\vk_2,\eta_{\ast}) } = - 2\, \textrm{Re} \lrsb{ i \, \int^{\eta_{\ast}}_{{ -\infty} } d\eta'\,
         \lrab{ Q(\vk_1,\eta_{\ast}) Q(\vk_2,\eta_{\ast}) Q(\vk_2,\eta_{\ast})\, H(\eta')  }
            } \,,
            }
where $\eta_{\ast}$ is the time when perturbation modes exit the
sound horizon, and $H$ is the Hamiltonian which can be read from Eq.
(\ref{s3_UV_fourier}) by noting that in the third-order $H_{(3)} = -
L_{(3)}$. Thus, for three-point interactions described by Eq.
(\ref{s3_UV_fourier}), we have \ea{{\label{3pf}}
    &\quad\;\lrab{ Q(\vk_1,\eta_{\ast}) Q(\vk_2,\eta_{\ast}) Q(\vk_2,\eta_{\ast})
    } \\
    &= (2\pi)^3 \delta^3(\vk_1 + \vk_2 + \vk_3) \, \textrm{Re} \lrsb{ \int_{-\infty}^{\eta_{\ast}} d\eta\,
        \frac{1}{a^5(\eta)H } e^{-i \sqrt{\frac{g_3}{\gamma }} \left(k_1^3+k_2^3+k_3^3\right) \int_{\eta}^{\eta_{\ast} } \frac{d\eta'}{a(\eta' )^2} }
        } \mathcal{S}(k_1,k_2,k_3) \,,
        }
where we have introduced the ``shape factor''
$\mathcal{S}(k_1,k_2,k_3)$  defined as
\ea{
    \mathcal{S}(k_1,k_2,k_3) &\equiv    \frac{ 3b_1}{2\gamma^3}  + \frac{b_3}{2\gamma^3}
\lrsb{ (\hat{\vk}_1 \cdot \hat{\vk}_2)^2 + (\hat{\vk}_2 \cdot
\hat{\vk}_3)^2 + (\hat{\vk}_3 \cdot \hat{\vk}_1)^2 }
         \\
         &\qquad\qquad + \frac{b_2}{4 \gamma ^2 g_3} \lrsb{ \frac{  k_1^2 \lrp{ k_2+k_3 } +k_2^2 \lrp{ k_3+k_1 } + k_3^2 \lrp{ k_1+k_2 } }{ k_1 k_2 k_3 }
        } \,.
}

For power law expansion $a(\eta) \propto |\eta|^{p}$ ($p\neq 0$),
the time-integral in Eq. (\ref{3pf}) can be evaluated exactly, and the
three-point correlation function reads, \eq{
    \lrab{ Q(\vk_1,\eta_{\ast}) Q(\vk_2,\eta_{\ast})
        Q(\vk_2,\eta_{\ast}) } \\
        = (2\pi)^3 \delta^3(\vk_1 + \vk_2 +
        \vk_3) B(k_1,k_2,k_3) \,,
} with
    \eq{
        B(k_1,k_2,k_3) =  \frac{1-2p}{p}\frac{ \gamma\, \mathcal{S}(k_1,k_2,k_3) }{g_3
    \left(k_1^3+k_2^3+k_3^3\right)^2}  \,,
    }
which is the so-called bispectrum.

\subsection{Non-linear parameter $f_{\textrm{NL}}$}

In practice, it is convenient to introduce non-linear
parameters to characterize the non-gaussianities. The dimensionless
non-linear parameter $f_{\textrm{NL}}$ from the three-point
correlation function is defined as
    \eq{
        B(k_1,k_2,k_3) \equiv \frac{6}{5}
        f_{\textrm{NL}}(k_1,k_2,k_3) \lrsb{ P(k_1)P(k_2) + P(k_2)P(k_3) + P(k_3)P(k_1)
        } \,,
    }
where the power spectrum $P(k)$ is given in Eq. (\ref{powersp}). Note
that although dimensionless, $f_{\textrm{NL}}$ is in general
$k$-dependent. A straightforward calculation gives
    \ea{
        &\quad\; f_{\textrm{NL}}(k_1,k_2,k_3) =
        \frac{10}{3}\frac{\gamma^2(1-2p)}{p} \frac{ k_1^3 k_2^3k_3^3 }{(k_1^3 + k_2^3 +
        k_3^3)^3} \mathcal{S}(k_1,k_2,k_3) \,.
    }

In the equilateral limit ($k_1 \approx k_2 \approx k_3$),
    \eq{
        f_{\textrm{NL}}^{\textrm{equil}} \approx \frac{5 (1-2 p) (\lambda-1 ) (91+3 \lambda  (-55+36 \lambda ))}{72 p\sqrt{6 \alpha} (3 \lambda -1)^{3/2} (7+3 \lambda  (-4+3 \lambda
        ))} \sim \mathcal{O}(1) \,,
    }
while in the squeezed limit ($k_1 \ll k_2 \approx k_3$),
    \eq{
        f_{\textrm{NL}}^{\textrm{local}} \approx \frac{5 (1-2 p) (\lambda -1)}{8 p\sqrt{6 \alpha}  (3 \lambda -1)^{3/2}} \lrp{\frac{k_1}{k_2}}^2 \ll 1 \,.
    }
Thus, in the UV limit, we find that the equilateral-type non-gaussianity
is roughly $\sim \mathcal{O}(1)$, while the local-type
non-gaussianity is very small. This is not surprising, since in
Ho\v{r}ava-Lifshitz gravity the scalar-field perturbation in the UV
limit is naturally scale-invariant. Thus, no slow-roll condition is
needed to guarantee the exponential expansion of the Universe. On
the other hand, it is well known that in standard slow-roll
inflationary models in GR non-gaussianities are suppressed by
slow-roll parameters \cite{Maldacena:2002vr}. However, in Ho\v{r}ava
gravity, no slow-roll parameters are needed. Thus, one may expect the
slow-roll suppressed non-gaussianity to be of the order of
unity. However, in our simplest scalar-field model with the action given by Eq.
(\ref{scalar_action}) and Eq. (\ref{scalar_f}), the temporal kinetic
term is canonical, and in general there is no enhancement of the
non-gaussianity by non-canonical kinetic terms as in the K-inflation
or DBI-inflation models \cite{Seery:2005wm,Chen:2006nt,Huang:2006eha,Arroja:2008ga}.
Moreover, the scalar-field action Eq. (\ref{scalar_action}) is mostly
``derivative-coupled'', thus the local-type non-gaussianity (which
characterizes the local couplings of the perturbations in the real space) is
small, as expected.

\section{Conclusion}

In this work, we investigated the cosmological perturbation theory
in Ho\v{r}ava-Lifshitz gravity and the non-gaussianities from the bispectrum. The most
interesting feature of Ho\v{r}ava gravity is that in the UV limit
the scalar perturbation is essentially scale-invariant, ignoring the
details of the expansion of the Universe. Moreover, together with
the fact that the speed of light in the UV limit diverges, there is
no need to assume exponential expansion of the early Universe, or
the usual scalar-field driven slow-roll inflation. In particular, the
slow-roll conditions are not necessary. Thus, one may expect that in
the absence of slow-roll conditions, the non-gaussianities can
become large. We calculated the three-point correlation function of the scalar
perturbation and found that the equilateral-type non-gaussianities
are of the order of unity due to the absence of the slow-roll-type
conditions, while the local-type non-gaussianities remain small, as
in the usual single field inflation in GR.

We focused on the scalar-field perturbation in the
Ho\v{r}ava-Lifshitz theory, neglecting the spatial metric
perturbations. However, the latter is obviously the most important
and interesting part in the Ho\v{r}ava theory. Since in the Ho\v{r}ava
theory the dynamical degrees of freedom in the spatial metric part are
3, especially, there is one additional scalar degree of freedom. It
is important to investigate the property of this additional degree
of freedom. Moreover, in this work, we only investigated the behavior
of the perturbation in the UV limit at $z=3$, while it is interesting to
study the full equations of motion, especially in the interpolating
region between UV and IR. We considered the scalar field with
canonical temporal kinetic term. However, one could expect
enhancement of non-gaussianities if more general kinetic
terms are considered. The Ho\v{r}ava-Lifshitz theory, although originating from a
renormalizable quantum gravity in 4 dimensions, may be
a potential competitor to the standard inflation theory and deserves
further investigation.

\bigskip

\textbf{Acknowledgements}

I would like to thank Robert Brandenberger, Miao Li, Chun-Shan Lin,
Yan Liu, Yi Pang, Ya-Wen Sun, Yi Wang, Jian-Feng Wu, Yushu Song,
Gang Yang and Yang Zhou for valuable discussions and comments. 
This work was partly supported by the Chinese National Youth Thousand Talents Program (71000-41180003).

\appendix

\section{Gauge Transformation, Gauge Choice and Gauge-invariant Variables}

In this Appendix, we discuss the problem of gauge transformation and
gauge choice in the non-relativistic Ho\v{r}ava-Lifshitz theory. The
essential point is that in the case of GR we have a larger set of
gauge transformations which we can use to choose a gauge. However, in the
Ho\v{r}ava-Lifshitz theory, the full diffeomorphism with general
coordinate invariance is restricted to a subset, i.e. the
so-called ``foliation-preserving diffeomorphism'', and thus gives us
less gauge modes but more physical modes.

In this work, we focus on the scalar-type perturbation. The scalar
part of coordinate transformations is : $\delta \eta = \xi^0$,
$\delta x^i = \rd^i \chi$, with
    \ea{
         \xi^0 &=\xi^0(\eta) \,,\\
        \chi &= \chi (\eta,x^i)
    }
and the scalar-type ``space-time metric'' perturbations
    \ea{
        g_{ij}  &= a^2 \lrsb{ (1-2\psi) \delta_{ij} + \rd_i\rd_j E }
        \\
        N &= 1 +2\phi \\
        N_i &= \rd_i B \,.
    }
The essential difference from GR is that in the
Ho\v{r}ava theory $\xi^0$ is a function of time $\eta$ only.

The gauge transformations is as usual:
    \ea{
        \Delta \phi &= -\frac{1}{a} \lrp{a \xi^0}' \\
        \Delta \psi &= \xi^0 \frac{a'}{a} \\
        \Delta B &= \xi^0 - \lrp{\frac{\chi}{a^2}}'\\
        \Delta E &= -\frac{\chi}{a^2}
    }

Due to the fact that $\xi^0$ is a function of $\eta$ only, the gauge
choice is fairly restricted. In particular, (or unfortunately), two
familiar gauge-choices --- the ``longitudinal gauge'' and ``spatially-flat
gauge'' --- are not allowed in the Ho\v{r}ava theory. We can use
$\chi= \chi(\eta,x^i)$ freely to set $E=0$, while in general we
cannot use $\xi^0$ to set $B=0$ and get the longitudinal gauge, or to set
$\psi=0$ and get the spatially-flat gauge.

However, there is still a possible gauge choice in the Ho\v{r}ava theory.
First, as we have mentioned, we can choose
    \[
        \chi = a^2 E
    \]
to get $\tilde{E}=0$. This leaves the question: are we
able to choose another gauge condition to set one of $\phi$, $\psi$
and $B$ to 0? In Ho\v{r}ava's original formulation of the theory,
$N=N(t)$ is assumed to be a function of time only. In this
case we can choose the proper value of $\xi^0$ to set the fluctuation
of $N$ to zero, i.e. $\phi=0$. Thus, after gauge transformation with
    \[
        \xi^0 = \frac{1}{a} \lrp{ \int d\eta\, a \phi(\eta) +c }\,,
    \]
we get $\tilde{\phi} = 0$. This does not determine  time-slicing
unambiguously, but we are left with time reparametrization. If we
relax the restriction that $N$ has to be a function of time only,
then there is no gauge condition we can choose. In this case,
we are left with 3 non-vanishing variables $\phi$, $\psi$ and $B$.

The two Bardeen potentials
    \ea{
        \Psi &= \psi - \frac{a'}{a}\lrp{B-E'} \\
        \Phi &= \phi + \frac{1}{a }\lrp{ a(B-E') }'
    }
are still gauge-invariant variables in Ho\v{r}ava gravity. This is also
because the foliation-preserving diffeormorphism is a subset of the
full symmetry in GR. Note that there is an infinite number of
gauge-invariant variables, for example, combining $\Phi$ and $\Psi$
gives another useful gauge-invariant variable
    \eq{
        \tilde{\Phi} = \phi + \frac{1}{a} \lrp{ \frac{\psi}{H}}' \,.
    }

For the ``space-time'' scalar field $\phi$ (now ``scalar'' means
invariant under ``foliation-preserving'' diffeomorphism), if we
assume that the background value is homogeneous $\phi_0 = \phi_0(\eta)$,
the gauge transformation for the scalar fluctuation is as usual
    \eq{
        \Delta \lrp{ \delta\phi} = - \xi^0 \phi'_0 \,.
    }
The gauge-invariant variable for $\delta\phi$ is as in GR:
    \eq{
        Q \equiv \delta\phi + \phi'_0 \lrp{B-E'} \,.
    }

%

\end{document}